\documentclass[10pt,pra,twocolumn,superscriptaddress]{revtex4-1}
\usepackage{graphicx}
\usepackage{amssymb}
\usepackage{times}
\usepackage{amsmath}
\usepackage{amsthm}
\usepackage{color}

\newcommand{\bra}[1]{\langle#1|}
\newcommand{\ket}[1]{|#1\rangle}
\newcommand{\tr}[0]{\mathrm{Tr}}

\newcommand{\SD}[0]{\mathrm{SD}}
\newcommand{\rhocode[1]}{\rho^{(#1)}}
\newcommand{\imaxc}[0]{I^{\textrm{max}}_c}

\begin{document}

\title{How Discord underlies the Noise Resilience of Quantum Illumination}

\author{Christian Weedbrook}
\affiliation{Department of Physics, University of Toronto, Toronto, M5S 3G4, Canada}

\author{Stefano Pirandola}\affiliation{Department of Computer Science, University of York, York YO10 5GH, United Kingdom}

\author{Jayne Thompson}\affiliation{Centre for Quantum Technologies, National University of Singapore, Singapore}

\author{Vlatko Vedral}\affiliation{Atomic and Laser Physics, Clarendon Laboratory,
University of Oxford, Parks Road, Oxford OX1 3PU, United Kingdom}
\affiliation{Centre for Quantum Technologies, National University of Singapore, Singapore}

\author{Mile Gu}
\email{cqtmileg@nus.edu.sg}
\affiliation{Center for Quantum Information, Institute for Interdisciplinary Information Sciences, Tsinghua University, Beijing, China}
\affiliation{Centre for Quantum Technologies, National University of Singapore, Singapore}
\affiliation{School of Mathematical and Physical Sciences, Nanyang Technological University, Singapore}

\date{\today}

\begin{abstract}The benefits of entanglement can outlast entanglement itself. In quantum illumination, entanglement is employed to better detect reflecting objects in environments so noisy that all entanglement is destroyed. Here, we show that quantum discord - a more resilient form of quantum correlations - explains the resilience of quantum illumination. We introduce a quantitative relation between the performance gain in quantum illumination and the amount of discord used to encode information about the presence or absence of a reflecting object. This highlights discord’s role preserving the benefits of entanglement in entanglement breaking noise.
\end{abstract}


\maketitle

\section{Introduction}

Quantum
illumination~\cite{Lloyd2008,Tan2008,Weedbrook2011,Shapiro2009,Guha2009,Barzanjeh2015}
offers a radical departure from conventional quantum
protocols~\cite{Nielsen2000,Wilde2013}. Most quantum technologies
require fragile entangling correlations to be preserved, whereas
quantum illumination operates in extremely-adverse environments
with entanglement-breaking noise~\cite{Sacchi2005,Piani2009}.
Specifically, quantum illumination aims to detect a low reflective
target basked in bright noise by probing it with one arm of an
entangled state. The protocol demonstrates significant improvement
over the use of conventional probes, even though the environmental
noise destroys all initial entanglement~\cite{Lloyd2008,Tan2008}.
This counter-intuitive phenomenon has been recently realized in
a series of experiments~\cite{Lopaeva2013,Zhang2003,Zhang2014}.%

The absence of entanglement, however, does not necessarily imply
classicality. Quantum protocols that operate with negligible
entanglement exist~\cite{Lanyon2008, Knill1998}, motivating the
search for quantum resources beyond entanglement. Quantum discord
is a prominent
candidate~\cite{Henderson2001,Ollivier2001,Modi2011}. Initially
proposed to isolate the `quantum' component of mutual information
between two physical systems, discord is conjectured to be a
potential quantum resource, responsible for the advantage of
certain quantum algorithms~\cite{Datta08}. While promising advances have
been made in understanding the operational significance of discord
~\cite{Zurek2003,Cavalcanti2011,boixo2011quantum,Madhok2011,Gu2012,Dakic2012,Adesso2013,girolami2013blind,Pirandola2014},
this remains a topic of significant debate. Contrary to
entanglement, which is difficult to synthesize, discord is
non-zero for almost every mixed state~\cite{Ferraro2010} and its
practical merit conflicts with the preconception that `quantum'
effects are fragile.

In this paper, we show that it is precisely the resilience of discord that explains the resilience of quantum illumination and highlight discord's role in preserving entanglement's benefits in quantum illumination. We first investigate what resources remain in illumination, after entanglement is broken by the environment. We then show that discord survives, and the quantum illumination makes use of this surviving discord to preserve information about the potential presence of a reflecting object that would otherwise be lost. We find that the amount of discord associated with sensing the target coincides exactly with the performance gain of quantum illumination over the best conventional technique.


\begin{figure}
\begin{center}
\includegraphics[width=7cm]{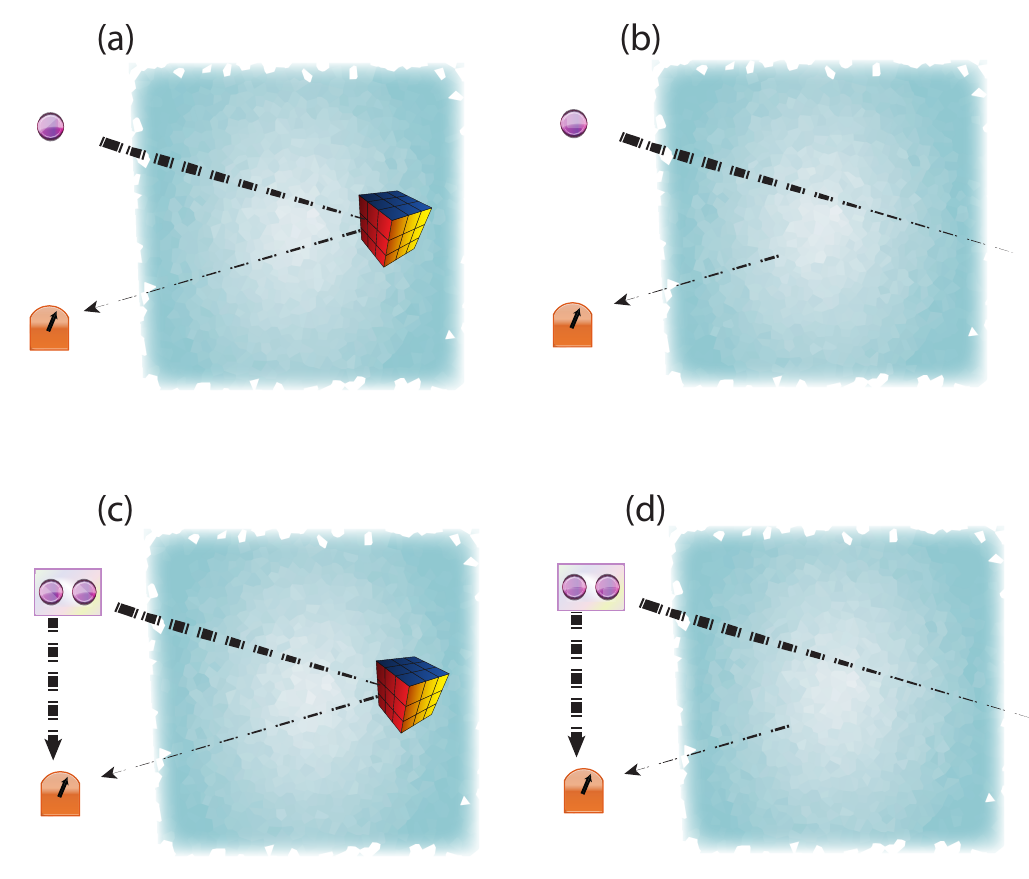}
\caption{\textbf{A premise of illumination}. In conventional
illumination, a single probe is sent into a noisy region to detect
the presence of a potential object. (a) If the object is
present, there is a small chance a reflected signal is detected;
otherwise (b) the probe is completely lost and Alice just sees only
random noise. In quantum illumination (c-d), Alice
prepares two maximally-entangled systems, one is kept (idler) and
the other sent for target detection (signal). The reflected signal and
idler are finally detected by a joint measurement. Surprisingly,
the use of an entangled source yields better performance, even though
entanglement fails to survive the return trip.}\label{fig:illum}
\label{fig1}
\end{center}
\end{figure}

\section{Framework} Illumination aims to discern whether
a weakly reflecting object is present or absent in a distant region
of intense noise (see Fig.~\ref{fig:illum}). This can be viewed as
a task in information retrieval. A distant region of space
contains a bit of information that dictates the presence ($x=0$)
or the absence ($x=1$) of the object. From this point
of view, the goal of illumination is to retrieve the value $x$ of
a random variable $X=\{x,p_{x}\}$ with binary alphabet
$x\in\{0,1\}$.


In the conventional approach, Alice probes the distant region with
a suitable quantum system (where suitable implies a system that
the reflector would potentially reflect), and monitors for a
potential reflection.  Let the probe be a $d$-dimensional quantum
system, i.e, a qudit, in a pure state $\Phi=\ket{\phi}\bra{\phi}$.
If the reflector is absent ($x = 1$), the entirety of $\Phi$ is
lost and Alice retrieves random environmental noise described by a
maximally mixed state $\rho_{E}=d^{-1}\mathbb{I}$ where
$\mathbb{I}$ is the identity operator. Otherwise ($x = 0$), the reflector
may reflect the object back at Alice; and the noise $\rho_E$ Alice observes is biased by the signal $\Phi$ with some small weighting $\eta \ll 1$. Thus,
probing the the reflector corresponds to encoding $x \in \{0,1\}$
into the output codewords
\begin{align}\nonumber
\rhocode[0]_c = \eta \Phi +(1-\eta) \rho_{E} \hspace{3mm} {\rm
and} \hspace{3mm} \rhocode[1]_c = \rho_{E}.
\end{align}
By detecting the reflected qudit, Alice has a limited ability to
distinguish these states and, therefore, to infer the value of
$x$.

In quantum illumination, Alice improves her strategy by resorting
to quantum correlations. She prepares a maximally-entangled state
$\Psi_{AB} = \ket{\psi}_{AB}\bra{\psi}$ of two qudits $A$ and $B$,
where $\ket{\psi}_{AB}=d^{-1} \sum_{k} \ket{k}_{A} \otimes
\ket{k}_{B}$, with $\{\ket{k}\}$ being an orthonormal basis. Then,
she probes the target with the signal system $A$ while retaining
the idler system $B$ in a quantum memory (or just a delay line in
experimental settings). Now we have encoded the value of $X$ via two codewords
\begin{align}\nonumber
\rho^{(0)}_{AB} = \eta \Psi_{AB} + (1-\eta) (\rho_{E} \otimes
\rho_{B}) \hspace{3mm} {\rm and} \hspace{3mm} \rho^{(1)}_{AB} =
\rho_E \otimes \rho_B,
\end{align}
where $\rho_B = \tr_A(\Psi_{AB})$ represents the reduced state of the idler if the signal is completely lost.

In either approach, Alice ends up in possession of one of two
potential codewords, $\rho^{(0)}$ or $\rho^{(1)}$, depending on
$x$. The better Alice can discriminate between these codewords,
the more information she can access about $x$. Quantum illumination thus
outperforms its conventional counterpart when it is easier to
distinguish $\rho^{(0)}_{AB}$ from $\rho^{(1)}_{AB}$, than $\rho^{(0)}_c$ from
$\rho^{(1)}_c$, for any conventional input $\Phi$.

To capture this quantity mathematically, consider first the general scenario where information about $X$ is encoded within a quantum system $S$; such that $S$ takes on the value $\rho^{(x)}$ when $X = x$. Let this encoding be captured by the ensemble $\varepsilon = \{p_x, \rho^{(x)}\}$, and $I$ be the amount of information about $X$ that Alice can access when given $\varepsilon$. That is, Alice is challenged to announce an estimate of $x$, $x_{\mathrm{est}}$, governed by random variable $X_{est}$. Her performance is dictated by the maximum $I(X,X_{\mathrm{est}})$ Alice can achieve, when supplied with $\rho^{(x)}$.

To evaluate $I$, observe that in order to retrieve information about $x$, Alice must measure some general positive operator value measurement (POVM) $\mathcal{M}$ on $S$, whose output defines another random variable $K^{\mathcal{M}}$ that is used by Alice to generate $X_{est}$. Alice's optimal performance thus aligns with the mutual information between $X$ and the measured output $K^{\mathcal{M}}$, when maximized over all possible measurements $\mathcal{M}$,\begin{equation}\label{Eq. Holevo}
I = \max_{\mathcal{M}}I(X,K^\mathcal{M}) = I_{acc}(\varepsilon)
\end{equation}
where $I(X,K^{\mathcal{M}}) = H(X) - H(X|K^\mathcal{M})$ and $I_{acc}(\varepsilon)$ denote's Alice's accessible information about $X$ with respect to $\varepsilon$. When $X$ is uniformly distributed, $I = \SD(\rho^{(0)},\rho^{(1)})$, where $\SD(\rho^{(0)},\rho^{(1)})$ is a well studied distinguishably measure known as the the Shannon distinguishably of $\rhocode[0]$ and $\rhocode[1]$.

\begin{figure}
\begin{center}
\includegraphics[width=8cm]{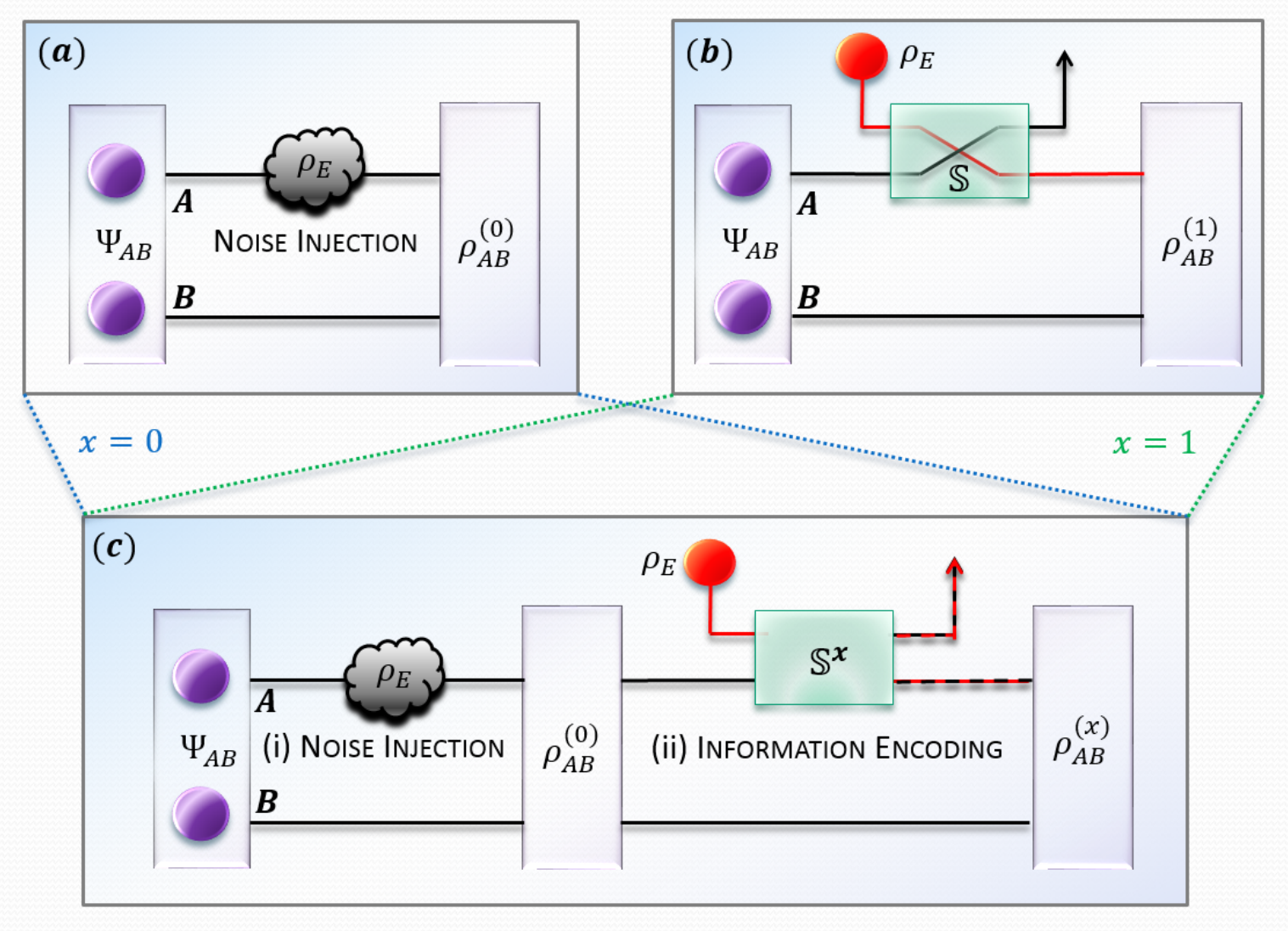}
\caption{\textbf{Operationally Equivalent Circuit Model of Illumination}. Quantum illumination can be understood in the quantum circuit picture. Let (a) and (b) denote respectively the operationally equivalent circuits for the presence ($x = 0$) and absence ($x = 1$) of the reflecting object. (a) is modeled by mixing $\Psi_{AB}$ with environmental noise $\rho_E$, resulting in state $\rho_{AB}^{(0)}$. In (b), the complete loss of signal is represented by a SWAP operation $\mathbb{S}$ between the signal system and environment. In either scenario, the resulting quantum channel is entanglement-breaking. The two scenarios can combine into a single circuit (c), composed of two sequential stages. In stage (i), environmental noise is injected into the signal arm by mixing in $\rho_E$, resulting in the state $\rho_{AB}^{(0)}$. In (ii), the presence or
absence of the target can be modeled as encoding the binary variable $X$ onto $\rho_{AB}$ by applying $\mathbb{S}^x$, where $\mathbb{S}^0=\mathbb{I}$ is the identity.}\label{Discord_schematic}
\end{center}
\end{figure}

Applying this result, the performance of quantum illumination is then given by $I_q = I_{acc}(\varepsilon_q)$, where $\varepsilon_q = \{p_x, \rhocode[x]_{AB}\}$. On the other hand, the optimal performance achievable in conventional illumination is provided by maximizing the accessible information with respect to the ensemble $\varepsilon_c = \{p_x, \rhocode[x]_c\}$ over all input states $\Phi$, i.e., $\imaxc = \max_{\Phi}I_{acc}(\varepsilon_c)$. The difference $\Delta I=I_q - \imaxc$ thus quantifies the advantage of quantum illumination - in terms of the amount of extra information Alice can gain about $x$ in a single trial. In the case of uniform $X$, $\Delta I = \SD(\rhocode[0]_{AB},\rhocode[1]_{AB}) - \max_{\Phi}\SD(\rhocode[0]_c,\rhocode[1]_c)$ is reduced to the gain in Shannon distinguishingly between codewords, when quantum methods are adopted over best conventional probes. While these quantities are generally very difficult to compute, the commutativity of the codewords makes the problem tractable for the special case of illumination (see appendix).

We note that several other methods to characterize the benefits of quantum illumination exist in literature. The quantifier introduced by Lloyd~\cite{Lloyd2008}, for example, is based on the probability of guessing $x$ correctly. The performance measures are closely related: Knowledge of one bounds the other from both above and below, and the scaling properties of the two measures coincide~\cite{Fuchs1999}. In using information theoretic quantifiers of distinguishability, we have followed an approach similar to that of quantum reading~\cite{Pirandola2011,pirandola2011quantum,dallarno2012}, where the mutual information
was used to better characterize the optimal readout of a classical memory.

\textbf{Noise Resilience}. The distinguishing feature of quantum illumination is that it exhibits a performance advantage even in scenarios where $\eta \ll 1$, and $\rho_E$ is completely mixed. This counters conventional intuition; the intense noise implies that $\rho^{(0)}_{AB}$ and $\rho^{(1)}_{AB}$ are
both highly entropic and completely separable, despite the use of a maximally entangled probe $\Psi_{AB}$. This peculiarity is highlighted when we recast quantum illumination into a functionally equivalent quantum circuit, where the action of the noise is separated from that of the reflecting object (see Fig.~\ref{Discord_schematic}). Irrespective of whether the reflector is present, the noise
decoheres Alice's input $\Psi_{AB}$ into the separable state
$\rhocode[0]_{AB}$ (cf. Fig.~\ref{Discord_schematic}.c). Now the
presence or absence of the target, i.e., the value $x$ of the random
variable $X$, is encoded into the state by applying the operator
$\mathbb{S}^{x}$ to the signal system, with $\mathbb{S}$ being the
swap operator between the signal and environment (cf.
Fig.~\ref{Discord_schematic}.b).

This viewpoint suggests that there must still exist some form of `quantumness' after noise injection; that is, we expect some form of quantum correlations to survive in the separable state $\rho^{(0)}_{AB}$ and that these correlations are related to quantum illumination's superior performance. Here, we demonstrate a direct relation between the discord remaining in $\rho^{(0)}_{AB}$ and the performance advantage in illumination, $\Delta I$.

\section{The Role of Discord} Formally, the discord of the
signal-idler system, denoted as $\delta(A|B)$, quantifies the
discrepancy between two types of correlations~\cite{Modi2011}. The
first type is the quantum mutual information $I(A,B)$ which
accounts for the total correlations between the two systems $A$
and $B$. The second type, denoted by $J(A|B)$, quantifies the
classical correlations and equals the maximal entropic reduction
of system $A$ under positive operator-value measure (POVM)
measurements $\{\Pi_b\}$ on system $B$. Explicitly, this is
defined by optimizing over all POVMs as $J(A|B) = S(A) -
\min_{\{\Pi_b\}}\sum p_b S(A|b)$, where $S(A)$ is the von Neumann
entropy of system $A$ and $S(A|b)$ is the entropy of system $A$
given the outcome $b$, achieved with probability $p_b$. The discord $\delta(A|B)=I(A,B)-J(A|B)$ between $A$ and $B$ captures the discrepancy between the two measures.

As aforementioned, quantum illumination can operate
when $\rho^{(0)}_{AB}$ (the state that is responsible for
sensing the target according to the equivalent quantum circuit in
Fig.~\ref{Discord_schematic}) - contains discord but no entanglement. In order to convince ourselves that this is more than just coincidental, we need to establish a quantitative relation between the discord that persists after noise injection, and the quantum advantage $\Delta I$. To do this, we draw inspiration from the concept of `discord consumption' that was used to highlight how discord can be interpreted as a resource that can be accessed via coherent interactions.

\begin{figure}
\begin{center}
\includegraphics[width=10cm]{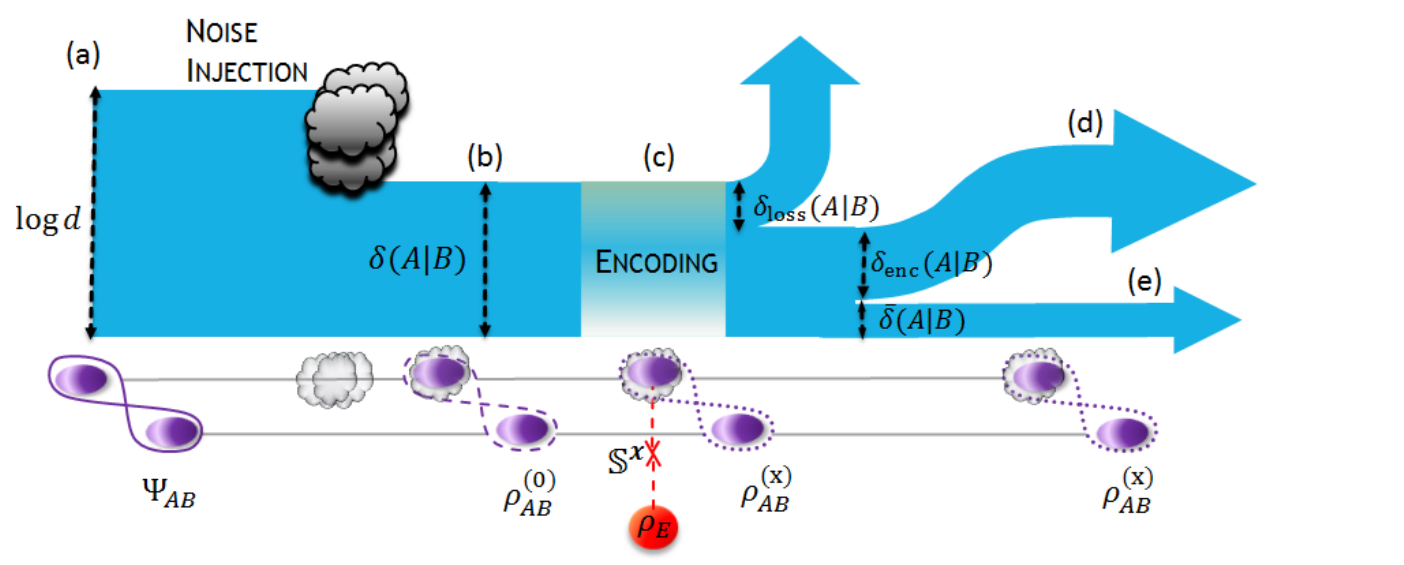}
\caption{\textbf{Discord flow in Illumination}. (a) Quantum illumination begins with a maximally entangled state with a discord value of $\log d$. (b) The injection of noise destroys all entanglement within the system producing a noisy state $\rho_{AB}^{(0)}$ which retains $\delta(A|B)$ units of discord. This represents the resources we effectively have available for sensing the target. (c) During the encoding stage, the swap gate $\mathbb{S}$ is applied with probability $p_1$, which leaks an average of  $\delta_{loss} = p_1\delta(A|B)$ units of discord into the environment, leaving $\delta(A|B) - \delta_{loss} = p_0\delta(A|B)$ units of discord available for encoding the value of $X$. (d) The encoding protocol then splits off $\delta_{enc}(A|B) = p_0\delta(A|B) - \overline{\delta}(A|B)$ to store extra information about $x$. (e) The remaining $\overline{\delta}(A|B)$ units of discord are not expended during the encoding protocol, and could be used to encode other information at some stage in the future.
\label{Discord consumption}}
\end{center}
\end{figure}

\textbf{Discord Expenditure.} Consider first a related scenario where Alice begins with a bipartite quantum state $\rho_{AB}$ with discord $\delta(A|B)$ as a resource. Alice encodes some $x$, governed by random variable $X$, by applying an $x$-dependent local unitary operation, $U_A^{(x)}$, on $A$; resulting in codewords $\rho_{AB}^{(x)}$. To a third party unaware of which $x$ is encoded, the resulting state is $\bar{\rho}_{AB} = \sum_k p_k \rho_{AB}^{(x)}$. Let $\bar{\delta}(A|B)$ be the discord of state $\bar{\rho}_{AB}$. The difference $\delta_\mathrm{enc}(A|B) = \delta(A|B) - \bar{\delta}(A|B)$ represents the reduction in discord from the perspective of a third party Bob, who is unaware of which $x$ was selected. In prior literature, this is regarded as the amount of discord consumed to encode $X$, or alternatively, the amount of information about $X$ that is encoded within discord correlations~\cite{Gu2012}.

In the above scenario, the transformation from initial resources to codewords used only local unitary operators on $A$. The discord of every individual codeword coincided with the discord of the original resource. Thus, no discord was lost to the environment during the encoding process. 
In illumination, this is no longer the case and we need to account for this extra loss as outlined by Fig. \ref{Discord consumption}. We make the following observations:
\begin{enumerate}
\item The amount of discord between signal and idler after noise injection is $\delta(A|B)$. This can be regarded as the amount of discorded resources we have prior to encoding.
\item The amount of discord after sensing the object is $\bar{\delta}(A|B)$ (for someone who does not know the value of $X$).
\item If a particular codeword $\rho_{AB}^{x}$ has discord $\delta^{x}(A|B) < \delta(A|B)$, then the encoding of $x$ is not discord preserving. In this case, we lose $\delta_{loss}^{x} = \delta(A|B) - \delta^{x}(A|B)$ units of discord. This discord is not used to encode $x$.
\item The average loss is then given by $\delta_{loss} = \sum_x p_x \delta_{loss}^{x}$.
 \end{enumerate}
In illumination $\delta_{loss} = p_1 \delta(A|B)$ as all $\delta(A|B)$ units of discord are lost if the reflecting object is absent. Factoring in this loss, we see that the amount of discord that is actually used to encode $x$ is given by
\begin{align}
\delta_{\mathrm{enc}}(A|B) &= \delta(A|B) - \delta_{loss}  - \bar{\delta}(A|B) \nonumber\\
&=  p_0 \delta(A|B) - \bar{\delta}(A|B).
\end{align}
This generalizes the concept of discord expended to encode the variable $x$ to the case of illumination. We can see that the only difference between this and the case of unitary encodings is the extra factor of $p_0$, representing that in illumination, only $p_0$ of the discorded resources before encoding are useful. Meanwhile, it shares the property that $\delta_{\mathrm{enc}}(A|B) \leq  \delta(A|B) - \delta_{loss} \leq \delta(A|B)$. The amount of discord associated with encoding $x$ is always abounded above by the amount of discord resources initially available. It is also interesting to note that $\delta_{\mathrm{enc}}(A|B) = \sum p_x \delta^{x}(A|B) - \bar{\delta}(A|B)$. That is, $\delta_{\mathrm{enc}}(A|B)$, can also be interpreted the gain in discord between signal and when someone learns the value of $x$.

\textbf{Relation to the Quantum Advantage.} The advantage of quantum illumination \emph{coincides exactly} with the discord expended for encoding $x$, that is
\begin{equation}\label{eq: final bound1}
\Delta I = \delta_{\mathrm{enc}}(A|B).
\end{equation}
The key idea behind our argument is as follows: We introduce an additional constraint to the quantum illumination protocol and show that
\begin{itemize}
\item[(i)] The optimal performance of quantum illumination, subject to this constraint, $I'_c$, coincides with the best performance using conventional illumination $I'_c = \imaxc$.
\item[(ii)] The loss in performance in enforcing this constraint over quantum illumination is $I_q - I'_c = \delta_{\mathrm{enc}}(A|B)$.
\end{itemize}
Specifically, the constraint imposes a specific measurement procedure Alice must use to extract $x$ upon receipt of $\rhocode[x]_{AB}$. Instead of allowing for arbitrary measurements, she is required to first make a local measurement on the idler $B$, followed by a local measurement on the signal $A$. (i) implies that this restricted procedure is operational equivalent to classical illumination, and (ii) implies that the loss of performance due to this restriction exactly coincides with the discord used to encode $x$. Together, the two statements imply the main result, i.e., $I_q - \imaxc = \delta_{\mathrm{enc}}(A|B)$. Details, including proofs of (i) and and (ii) are available in the appendix.

These results reveal why quantum illumination is advantageous, and how discord plays a role. In Gu et.al \cite{Gu2012}, it was established that information encoded within discorded correlations of two objects, $A$ and $B$, represents information that can only be extracted through coherent interactions between $A$ and $B$. Here, (i) indicates that that $\delta_{\mathrm{enc}}(A|B)$ represents information about $x$ that is encoded within discorded correlations, while (ii) demonstrates that quantum illumination derives it advantage by using coherent interactions between idler and probe to access this information.

As a result, quantum illumination gives an advantage $\Delta I
>0 $ only if the effective state $\rho_{AB}^{(0)}$ generated by the environment has non-zero discord, and the corresponding quantum advantage
$\Delta I$ is directly provided by the amount of discord
$\delta_{\mathrm{enc}}(A|B)$ associated with storing information about the presence and absence of the target. This identifies that discord plays a key role behind the resilience of quantum illumination, providing an extra resource in which information about the target is stored. While entanglement does not survive in quantum illumination, the survival of discord is essential for it to have any advantage over conventional illumination.

\section{A Simple Example} We illustrate the equivalence of
Eq.~(\ref{eq: final bound1}) in the case where signal and idler
are two-level quantum systems, i.e., qubits. The environment is flooded
with random qubits, such that $\rho_E = \mathbb{I}/2$. For
example, this may model the detection of a multi-faceted,
rotating, object in noise~\cite{Lloyd2008}.

The conventional approach probes the target with a pure state
$\ket{\phi}$, returning either $\rhocode[0]_c =
\eta\ket{\phi}\bra{\phi} + (1-\eta)\mathbb{I}/2$ or $\rhocode[1]_c
=  \mathbb{I}/2$ (any pure input state gives the same
performance). In quantum illumination, Alice instead probes the
target with one of arm of the Bell state $\ket{\psi} = (\ket{01} -
\ket{10})/\sqrt{2}$ or any other maximally-entangled state. This
results in codewords, $\rhocode[0]_{AB} = \eta
\ket{\psi}\bra{\psi}+ (1-\eta) \mathbb{I}/4$ and $\rhocode[1]_{AB}
= \mathbb{I}/4$. The corresponding performances of conventional
and quantum illumination, $\imaxc$ and $I_{q}$, respectively, are plotted versus the target reflectivity $\eta$ in Fig.~3.a for the case where $X$ is distributed unformly. The difference between
these curves (shaded region) quantifies the gain $\Delta I$ of
quantum illumination.

As we can see from Fig.~3.b, the state of the system after
noise, $\rhocode[0]_{AB}$, is always separable for sufficiently
small values of $\eta$. Nevertheless, $\rhocode[0]_{AB}$ contains
discord, part of which can be harnessed to store information about
$x$. In comparing Fig.~3.a with Fig.~3.b, we see that the amount
of discord expended for resolving the target
$\delta_{\mathrm{enc}}(A|B)$ coincides exactly with the advantage
$\Delta I$ of quantum illumination.

\begin{figure}
\begin{center}
\includegraphics[width=7cm]{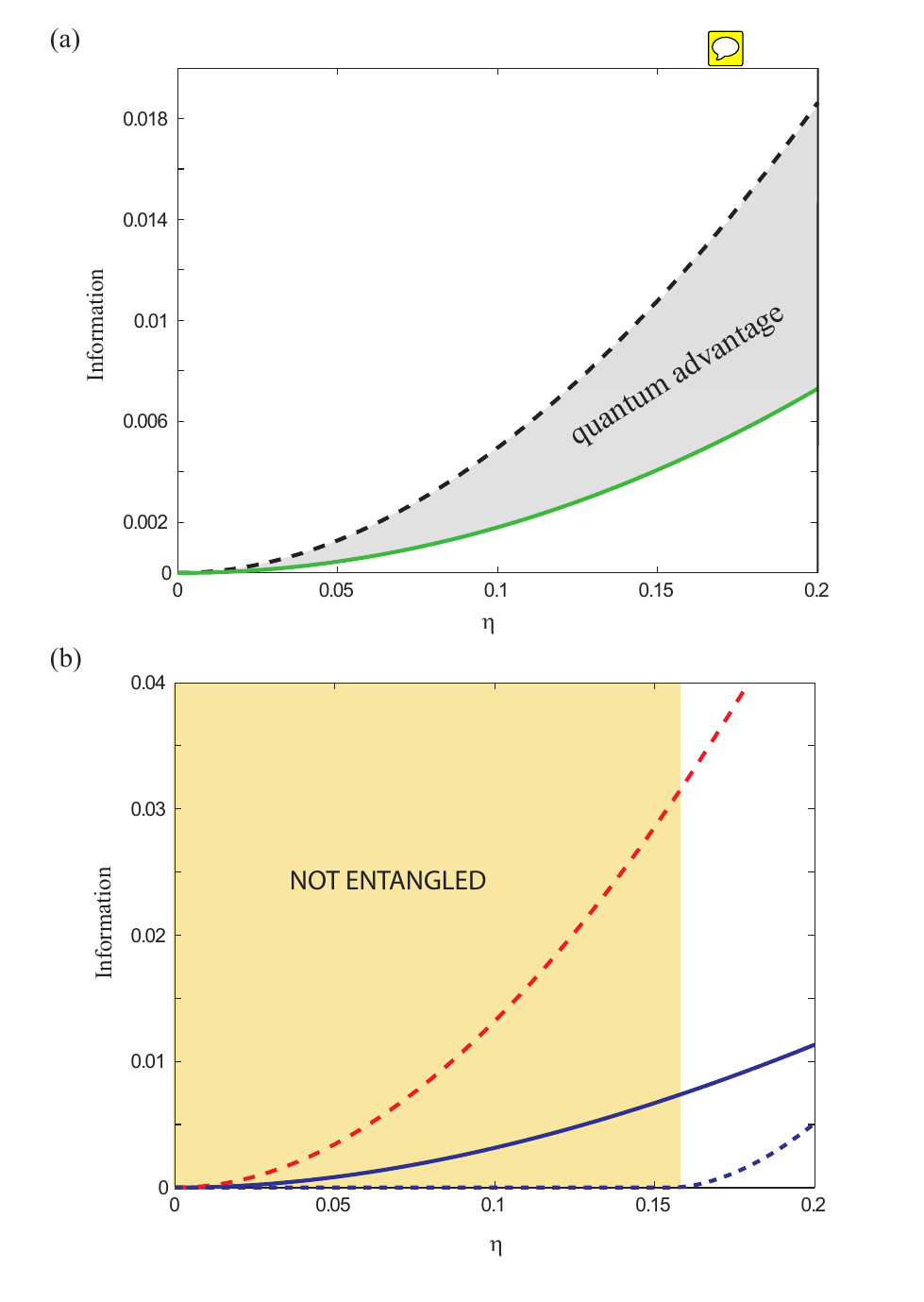}
\caption{\textbf{Advantage and discord in qubit illumination.} (a)
Quantum illumination (black dashed line) outperforms the best
conventional methods (green solid line) for all non-zero values of
$\eta$. (b) Entanglement of formation of the signal-idler system
is zero for $\eta < 0.15$ (blue dashed line). Discord, however, is
present for all values of $\eta$ (dashed red line), and the amount
of discord expended to encode $x$ (blue solid line), $\delta_{\mathrm{enc}}(A|B)$, coincides
exactly with the performance gain of quantum illumination, $\Delta I$, i.e., the shaded
region in (a).}\label{deltaI}
\end{center}
\end{figure}

\section{Discussion} In this paper, we have shown that discord underlies the resilience of quantum illumination in entanglement-breaking noise.  In such situations, discord can survive when entanglement does not. Quantum illumination exploits these surviving quantum correlations to encode extra information regarding the potential presence of a reflective object. The amount of discord used to encode this information is shown to \emph{coincide exactly} with the enhanced performance of quantum illumination, the equivalence holding for systems of arbitrary dimensions. This connection explains why the benefits of entanglement may survive entanglement-breaking noise, and helps establish discord's role in noise resilient quantum technology. The results in this manuscript are valid for general distributions of $X$. Thus our arguments apply to cases where one repeats the protocol multiple times to gain progressively more information about $X$. This can be modelled through Bayesian update, where the prior for $X$ is updated with each successive trial.

In deriving our results, we quantified both the discord between signal and idler, and the performance advantage of quantum illumination via entropic measures. There are, of course, many other ways to measure either quantify (e.g. geometric measures of discord, increased success probabilities to measure performance advantage) and one may well be able to obtain similar relations between suitable alternative measures. Indeed, considerations of other performance measures for illumination may well motivate new operational measures of discord, much as consideration of phase estimation motivated interferometric power~\cite{girolami2013blind}.

The techniques featured may also be generalized to related situations, such as encoding and communicating information when applying more general quantum operations in intense entanglement-breaking noise. This could lend insight to discord's role in cryptographic variants of illumination ~\cite{Zhang2003,Shapiro2009a,Pirandola2008}. Our analysis also have potential to generalize to the continuous variable regime, though the non-commutativity of the resulting codewords may make direct analytical approaches. If so though, it will complement concurrent approaches to understand continuous quantum illumination's operational advantage using mutual information~\cite{ragy2014}.

More generally, illumination belongs to a broader collection of protocols aimed to determine certain properties of unknown quantum channels, including quantum channel discrimination, quantum loss detection, and quantum metrology. In each of these protocols, numerical links between discord and performance have been proposed \cite{Datta08,Modi2011a,Invernizzi2011}. A similar approach to understanding how discord's role in preserving information in more general bipartite encodings could further formalize discord's influence in such scenarios, and lead to a unified, information theoretic understanding of how the benefits of entanglement survive when entanglement dies.

\acknowledgements We would like to thank Si-Hui Tan, Ping Koy Lam,
Syed Assad, Thomas Symul, Marco Piani, and Jing Yan Haw for
discussions. This work is supported by the NSERC, the Leverhulme
Trust, the Engineering and Physical Sciences Research Council
(EPSRC) Grant EP/L011298/1, the National Research Foundation, the
Ministry Education in Singapore Grant and the Academic Research
Fund Tier 3 MOE2012-T3-1-009, the National Basic Research Program
of China Grant 2011CBA00300, 2011CBA00302, the National Natural
Science Foundation of China Grant 11450110058, 61033001,
61361136003 and the John Templeton Foundation Grant 53914 {\em ``Occam's Quantum Mechanical Razor: Can Quantum theory admit the Simplest Understanding of Reality?''}.


\begin{thebibliography}{33}
\expandafter\ifx\csname natexlab\endcsname\relax\def\natexlab#1{#1}\fi
\expandafter\ifx\csname bibnamefont\endcsname\relax
  \def\bibnamefont#1{#1}\fi
\expandafter\ifx\csname bibfnamefont\endcsname\relax
  \def\bibfnamefont#1{#1}\fi
\expandafter\ifx\csname citenamefont\endcsname\relax
  \def\citenamefont#1{#1}\fi
\expandafter\ifx\csname url\endcsname\relax
  \def\url#1{\texttt{#1}}\fi
\expandafter\ifx\csname urlprefix\endcsname\relax\def\urlprefix{URL }\fi
\providecommand{\bibinfo}[2]{#2}
\providecommand{\eprint}[2][]{\url{#2}}

\bibitem[{\citenamefont{Lloyd}(2008)}]{Lloyd2008}
\bibinfo{author}{\bibfnamefont{S.}~\bibnamefont{Lloyd}},
  \bibinfo{journal}{Science} \textbf{\bibinfo{volume}{321}},
  \bibinfo{pages}{1463} (\bibinfo{year}{2008}).

\bibitem[{\citenamefont{Tan et~al.}(2008)\citenamefont{Tan, Erkmen,
  Giovannetti, Guha, Lloyd, Maccone, Pirandola, and Shapiro}}]{Tan2008}
\bibinfo{author}{\bibfnamefont{S.-H.} \bibnamefont{Tan}},
  \bibinfo{author}{\bibfnamefont{B.~I.} \bibnamefont{Erkmen}},
  \bibinfo{author}{\bibfnamefont{V.}~\bibnamefont{Giovannetti}},
  \bibinfo{author}{\bibfnamefont{S.}~\bibnamefont{Guha}},
  \bibinfo{author}{\bibfnamefont{S.}~\bibnamefont{Lloyd}},
  \bibinfo{author}{\bibfnamefont{L.}~\bibnamefont{Maccone}},
  \bibinfo{author}{\bibfnamefont{S.}~\bibnamefont{Pirandola}},
  \bibnamefont{and} \bibinfo{author}{\bibfnamefont{J.~H.}
  \bibnamefont{Shapiro}}, \bibinfo{journal}{Phys. Rev. Lett.}
  \textbf{\bibinfo{volume}{101}}, \bibinfo{pages}{253601}
  (\bibinfo{year}{2008}).

\bibitem[{\citenamefont{Weedbrook et~al.}(2011)\citenamefont{Weedbrook, Pirandola, Garc\'ia-Patr\'on, Cerf, Ralph, Shapiro, and
Lloyd}}]{Weedbrook2011}
\bibinfo{author}{\bibfnamefont{C.}~\bibnamefont{Weedbrook}},
\bibinfo{author}{\bibfnamefont{S.} \bibnamefont{Pirandola}},
\bibinfo{author}{\bibfnamefont{R.}~\bibnamefont{Garc\'ia-Patr\'on}},
\bibinfo{author}{\bibfnamefont{N.~J.} \bibnamefont{Cerf}},
\bibinfo{author}{\bibfnamefont{T.~C.} \bibnamefont{Ralph}},
\bibinfo{author}{\bibfnamefont{J.~H.} \bibnamefont{Shapiro}},
\bibnamefont{and}
\bibinfo{author}{\bibfnamefont{S.} \bibnamefont{Lloyd}},
\bibinfo{journal}{Rev. Mod. Phys.} \textbf{\bibinfo{volume}{84}},
\bibinfo{pages}{621} (\bibinfo{year}{2012}).


\bibitem[{\citenamefont{Shapiro and Lloyd}(2009)}]{Shapiro2009}
\bibinfo{author}{\bibfnamefont{J.~H.} \bibnamefont{Shapiro}} \bibnamefont{and}
  \bibinfo{author}{\bibfnamefont{S.}~\bibnamefont{Lloyd}},
  \bibinfo{journal}{New Journal of Physics} \textbf{\bibinfo{volume}{11}},
  \bibinfo{pages}{063045} (\bibinfo{year}{2009}).


\bibitem[{\citenamefont{Guha and Erkmen}(2009)}]{Guha2009}
\bibinfo{author}{\bibfnamefont{S.}~\bibnamefont{Guha}} \bibnamefont{and}
  \bibinfo{author}{\bibfnamefont{B.~I.} \bibnamefont{Erkmen}},
  \bibinfo{journal}{Phys. Rev. A} \textbf{\bibinfo{volume}{80}},
  \bibinfo{pages}{052310} (\bibinfo{year}{2009}).

  \bibitem[{\citenamefont{Barzanjeh et~al.}(2015)\citenamefont{Guha, Weedbrook, Pirandola, Vitali, Cerf, Shapiro}}]{Barzanjeh2015}
\bibinfo{author}{\bibfnamefont{S.}~\bibnamefont{Barzanjeh}},
\bibinfo{author}{\bibfnamefont{S.}~\bibnamefont{Guha}},
\bibinfo{author}{\bibfnamefont{C.} \bibnamefont{Weedbrook}},
\bibinfo{author}{\bibfnamefont{C.} \bibnamefont{Vitali}},
\bibinfo{author}{\bibfnamefont{J.~H.} \bibnamefont{Shapiro}},
\bibnamefont{and}
\bibinfo{author}{\bibfnamefont{S.} \bibnamefont{Pirandola}},
\bibinfo{journal}{Phys. Rev. Lett.} \textbf{\bibinfo{volume}{114}},
\bibinfo{pages}{080503} (\bibinfo{year}{2015}).

\bibitem[{\citenamefont{Nielsen and Chuang}(2010)}]{Nielsen2000}
\bibinfo{author}{\bibfnamefont{M.~A.} \bibnamefont{Nielsen}} \bibnamefont{and}
  \bibinfo{author}{\bibfnamefont{I.~L.} \bibnamefont{Chuang}},
  \emph{\bibinfo{title}{Quantum computation and quantum information}}
  (\bibinfo{publisher}{Cambridge university press, Cambridge}, \bibinfo{year}{2010}).

\bibitem[{\citenamefont{Wilde}(2013)}]{Wilde2013}
\bibinfo{author}{\bibfnamefont{M.~M.} \bibnamefont{Wilde}},
  \emph{\bibinfo{title}{Quantum Information Theory}}
  (\bibinfo{publisher}{Cambridge university press, Cambridge}, \bibinfo{year}{2013}).


\bibitem[{\citenamefont{Sacchi}(2005)}]{Sacchi2005}
\bibinfo{author}{\bibfnamefont{M.~F.} \bibnamefont{Sacchi}},
  \bibinfo{journal}{Phys. Rev. A} \textbf{\bibinfo{volume}{72}},
  \bibinfo{pages}{014305} (\bibinfo{year}{2005}).

\bibitem[{\citenamefont{Piani and Watrous}(2005)}]{Piani2009}
\bibinfo{author}{\bibfnamefont{M.} \bibnamefont{Piani}}
\bibnamefont{and}
\bibinfo{author}{\bibfnamefont{J.} \bibnamefont{Watrous}},
\bibinfo{journal}{Phys. Rev. Lett.} \textbf{\bibinfo{volume}{102}},
\bibinfo{pages}{250501} (\bibinfo{year}{2009}).


\bibitem[{\citenamefont{Lopaeva et~al.}(2013)\citenamefont{Lopaeva,
  Ruo~Berchera, Degiovanni, Olivares, Brida, and Genovese}}]{Lopaeva2013}
\bibinfo{author}{\bibfnamefont{E.~D.} \bibnamefont{Lopaeva}},
  \bibinfo{author}{\bibfnamefont{I.}~\bibnamefont{Ruo~Berchera}},
  \bibinfo{author}{\bibfnamefont{I.~P.} \bibnamefont{Degiovanni}},
  \bibinfo{author}{\bibfnamefont{S.}~\bibnamefont{Olivares}},
  \bibinfo{author}{\bibfnamefont{G.}~\bibnamefont{Brida}}, \bibnamefont{and}
  \bibinfo{author}{\bibfnamefont{M.}~\bibnamefont{Genovese}},
  \bibinfo{journal}{Phys. Rev. Lett.} \textbf{\bibinfo{volume}{110}},
  \bibinfo{pages}{153603} (\bibinfo{year}{2013}).

\bibitem[{\citenamefont{Zhang et~al.}(2013)\citenamefont{Zhang, Tengner, Zhong,
  Wong, and Shapiro}}]{Zhang2003}
\bibinfo{author}{\bibfnamefont{Z.}~\bibnamefont{Zhang}},
  \bibinfo{author}{\bibfnamefont{M.}~\bibnamefont{Tengner}},
  \bibinfo{author}{\bibfnamefont{T.}~\bibnamefont{Zhong}},
  \bibinfo{author}{\bibfnamefont{F.~N.~C.} \bibnamefont{Wong}},
  \bibnamefont{and} \bibinfo{author}{\bibfnamefont{J.~H.}
  \bibnamefont{Shapiro}}, \bibinfo{journal}{Phys. Rev. Lett.}
  \textbf{\bibinfo{volume}{111}}, \bibinfo{pages}{010501}
  (\bibinfo{year}{2013}).

\bibitem[{\citenamefont{Zhang et~al.}(2014)\citenamefont{Zhang, Mouradian, Wong, and Shapiro}}]{Zhang2014}
\bibinfo{author}{\bibfnamefont{Z.}~\bibnamefont{Zhang}},
  \bibinfo{author}{\bibfnamefont{S.}~\bibnamefont{Mouradian}},
  \bibinfo{author}{\bibfnamefont{F.~N.~C.} \bibnamefont{Wong}},
  \bibnamefont{and} \bibinfo{author}{\bibfnamefont{J.~H.}
  \bibnamefont{Shapiro}}, \bibinfo{journal}{Phys. Rev. Lett.} \textbf{\bibinfo{volume}{114}},
   \bibinfo{pages}{110506} (\bibinfo{year}{2015}{\natexlab{b}}).


\bibitem[{\citenamefont{Lanyon et~al.}(2008)\citenamefont{Lanyon, Barbieri,
  Almeida, and White}}]{Lanyon2008}
\bibinfo{author}{\bibfnamefont{B.~P.} \bibnamefont{Lanyon}},
  \bibinfo{author}{\bibfnamefont{M.}~\bibnamefont{Barbieri}},
  \bibinfo{author}{\bibfnamefont{M.~P.} \bibnamefont{Almeida}},
  \bibnamefont{and} \bibinfo{author}{\bibfnamefont{A.~G.} \bibnamefont{White}},
  \bibinfo{journal}{Phys. Rev. Lett.} \textbf{\bibinfo{volume}{101}},
  \bibinfo{pages}{200501} (\bibinfo{year}{2008}).

\bibitem[{\citenamefont{Knill and Laflamme}(1998)}]{Knill1998}
\bibinfo{author}{\bibfnamefont{E.}~\bibnamefont{Knill}} \bibnamefont{and}
  \bibinfo{author}{\bibfnamefont{R.}~\bibnamefont{Laflamme}},
  \bibinfo{journal}{Phys. Rev. Lett.} \textbf{\bibinfo{volume}{81}},
  \bibinfo{pages}{5672} (\bibinfo{year}{1998}).

\bibitem[{\citenamefont{Henderson and Vedral}(2001)}]{Henderson2001}
\bibinfo{author}{\bibfnamefont{L.}~\bibnamefont{Henderson}} \bibnamefont{and}
  \bibinfo{author}{\bibfnamefont{V.}~\bibnamefont{Vedral}},
  \bibinfo{journal}{J. Phys. A}
  \textbf{\bibinfo{volume}{34}}, \bibinfo{pages}{6899} (\bibinfo{year}{2001}).

\bibitem[{\citenamefont{Ollivier and Zurek}(2001)}]{Ollivier2001}
\bibinfo{author}{\bibfnamefont{H.}~\bibnamefont{Ollivier}} \bibnamefont{and}
  \bibinfo{author}{\bibfnamefont{W.~H.} \bibnamefont{Zurek}},
  \bibinfo{journal}{Phys. Rev. Lett.} \textbf{\bibinfo{volume}{88}},
  \bibinfo{pages}{017901} (\bibinfo{year}{2001}).

\bibitem[{\citenamefont{Modi et~al.}(2012)\citenamefont{Modi, Brodutch, Cable,
  Paterek, and Vedral}}]{Modi2011}
\bibinfo{author}{\bibfnamefont{K.}~\bibnamefont{Modi}},
  \bibinfo{author}{\bibfnamefont{A.}~\bibnamefont{Brodutch}},
  \bibinfo{author}{\bibfnamefont{H.}~\bibnamefont{Cable}},
  \bibinfo{author}{\bibfnamefont{T.}~\bibnamefont{Paterek}}, \bibnamefont{and}
  \bibinfo{author}{\bibfnamefont{V.}~\bibnamefont{Vedral}},
  \bibinfo{journal}{Rev. Mod. Phys.} \textbf{\bibinfo{volume}{84}},
  \bibinfo{pages}{1655} (\bibinfo{year}{2012}).

\bibitem[{\citenamefont{Datta et~al.}(2008)\citenamefont{Datta, Shaji, and
  Caves}}]{Datta08}
\bibinfo{author}{\bibfnamefont{A.}~\bibnamefont{Datta}},
  \bibinfo{author}{\bibfnamefont{A.}~\bibnamefont{Shaji}}, \bibnamefont{and}
  \bibinfo{author}{\bibfnamefont{C.~M.} \bibnamefont{Caves}},
  \bibinfo{journal}{Phys. Rev. Lett.} \textbf{\bibinfo{volume}{100}},
  \bibinfo{pages}{050502} (\bibinfo{year}{2008}).

\bibitem[{\citenamefont{Zurek}(2003)}]{Zurek2003}
\bibinfo{author}{\bibfnamefont{W.~H.} \bibnamefont{Zurek}},
  \bibinfo{journal}{Phys. Rev. A} \textbf{\bibinfo{volume}{67}},
  \bibinfo{pages}{012320} (\bibinfo{year}{2003}).

\bibitem[{\citenamefont{Cavalcanti et~al.}(2011)\citenamefont{Cavalcanti,
  Aolita, Boixo, Modi, Piani, and Winter}}]{Cavalcanti2011}
\bibinfo{author}{\bibfnamefont{D.}~\bibnamefont{Cavalcanti}},
  \bibinfo{author}{\bibfnamefont{L.}~\bibnamefont{Aolita}},
  \bibinfo{author}{\bibfnamefont{S.}~\bibnamefont{Boixo}},
  \bibinfo{author}{\bibfnamefont{K.}~\bibnamefont{Modi}},
  \bibinfo{author}{\bibfnamefont{M.}~\bibnamefont{Piani}}, \bibnamefont{and}
  \bibinfo{author}{\bibfnamefont{A.}~\bibnamefont{Winter}},
  \bibinfo{journal}{Phys. Rev. A} \textbf{\bibinfo{volume}{83}},
  \bibinfo{pages}{032324} (\bibinfo{year}{2011}).

\bibitem[{\citenamefont{Boixo et~al.}(2011)\citenamefont{Boixo, Aolita,
  Cavalcanti, Modi, and Winter}}]{boixo2011quantum}
\bibinfo{author}{\bibfnamefont{S.}~\bibnamefont{Boixo}},
  \bibinfo{author}{\bibfnamefont{L.}~\bibnamefont{Aolita}},
  \bibinfo{author}{\bibfnamefont{D.}~\bibnamefont{Cavalcanti}},
  \bibinfo{author}{\bibfnamefont{K.}~\bibnamefont{Modi}}, \bibnamefont{and}
  \bibinfo{author}{\bibfnamefont{A.}~\bibnamefont{Winter}},
  \bibinfo{journal}{International Journal of Quantum Information}
  \textbf{\bibinfo{volume}{9}}, \bibinfo{pages}{1643} (\bibinfo{year}{2011}).

\bibitem[{\citenamefont{Madhok and Datta}(2011)}]{Madhok2011}
\bibinfo{author}{\bibfnamefont{V.}~\bibnamefont{Madhok}} \bibnamefont{and}
  \bibinfo{author}{\bibfnamefont{A.}~\bibnamefont{Datta}},
  \bibinfo{journal}{Phys. Rev. A} \textbf{\bibinfo{volume}{83}},
  \bibinfo{pages}{032323} (\bibinfo{year}{2011}).

\bibitem[{\citenamefont{Gu et~al.}(2012)\citenamefont{Gu, Chrzanowski, Assad,
  Symul, Modi, Ralph, Vedral, and Lam}}]{Gu2012}
\bibinfo{author}{\bibfnamefont{M.}~\bibnamefont{Gu}},
  \bibinfo{author}{\bibfnamefont{H.~M.} \bibnamefont{Chrzanowski}},
  \bibinfo{author}{\bibfnamefont{S.~M.} \bibnamefont{Assad}},
  \bibinfo{author}{\bibfnamefont{T.}~\bibnamefont{Symul}},
  \bibinfo{author}{\bibfnamefont{K.}~\bibnamefont{Modi}},
  \bibinfo{author}{\bibfnamefont{T.~C.} \bibnamefont{Ralph}},
  \bibinfo{author}{\bibfnamefont{V.}~\bibnamefont{Vedral}}, \bibnamefont{and}
  \bibinfo{author}{\bibfnamefont{P.~K.} \bibnamefont{Lam}},
  \bibinfo{journal}{Nature Physics}  (\bibinfo{year}{2012}).


\bibitem[{\citenamefont{Daki{\'c} et~al.}(2012)\citenamefont{Daki{\'c}, Lipp,
  Ma, Ringbauer, Kropatschek, Barz, Paterek, Vedral, Zeilinger, Brukner
  et~al.}}]{Dakic2012}
\bibinfo{author}{\bibfnamefont{B.}~\bibnamefont{Daki{\'c}}},
  \bibinfo{author}{\bibfnamefont{Y.~O.} \bibnamefont{Lipp}},
  \bibinfo{author}{\bibfnamefont{X.}~\bibnamefont{Ma}},
  \bibinfo{author}{\bibfnamefont{M.}~\bibnamefont{Ringbauer}},
  \bibinfo{author}{\bibfnamefont{S.}~\bibnamefont{Kropatschek}},
  \bibinfo{author}{\bibfnamefont{S.}~\bibnamefont{Barz}},
  \bibinfo{author}{\bibfnamefont{T.}~\bibnamefont{Paterek}},
  \bibinfo{author}{\bibfnamefont{V.}~\bibnamefont{Vedral}},
  \bibinfo{author}{\bibfnamefont{A.}~\bibnamefont{Zeilinger}},
  \bibinfo{author}{\bibfnamefont{{\v{C}}.}~\bibnamefont{Brukner}},
  \bibnamefont{et~al.}, \bibinfo{journal}{Nature Physics}
  \textbf{\bibinfo{volume}{8}}, \bibinfo{pages}{666} (\bibinfo{year}{2012}).


\bibitem[{\citenamefont{Girolami
  et~al.}(2013{\natexlab{a}})\citenamefont{Girolami, Tufarelli, and
  Adesso}}]{Adesso2013}
\bibinfo{author}{\bibfnamefont{D.}~\bibnamefont{Girolami}},
  \bibinfo{author}{\bibfnamefont{T.}~\bibnamefont{Tufarelli}},
  \bibnamefont{and} \bibinfo{author}{\bibfnamefont{G.}~\bibnamefont{Adesso}},
  \bibinfo{journal}{Phys. Rev. Lett.} \textbf{\bibinfo{volume}{110}},
  \bibinfo{pages}{240402} (\bibinfo{year}{2013}{\natexlab{a}}).



\bibitem[{\citenamefont{Girolami
  et~al.}(2013{\natexlab{b}})\citenamefont{Girolami, Souza, Giovannetti,
  Tufarelli, Filgueiras, Sarthour, Soares-Pinto, Oliveira, and
  Adesso}}]{girolami2013blind}
\bibinfo{author}{\bibfnamefont{D.}~\bibnamefont{Girolami}},
  \bibinfo{author}{\bibfnamefont{A.~M.} \bibnamefont{Souza}},
  \bibinfo{author}{\bibfnamefont{V.}~\bibnamefont{Giovannetti}},
  \bibinfo{author}{\bibfnamefont{T.}~\bibnamefont{Tufarelli}},
  \bibinfo{author}{\bibfnamefont{J.~G.} \bibnamefont{Filgueiras}},
  \bibinfo{author}{\bibfnamefont{R.~S.} \bibnamefont{Sarthour}},
  \bibinfo{author}{\bibfnamefont{D.~O.} \bibnamefont{Soares-Pinto}},
  \bibinfo{author}{\bibfnamefont{I.~S.} \bibnamefont{Oliveira}},
  \bibnamefont{and} \bibinfo{author}{\bibfnamefont{G.}~\bibnamefont{Adesso}},
  \bibinfo{journal}{Phys. Rev. Lett} \textbf{\bibinfo{volume}{112}},
  \bibinfo{pages}{221} (\bibinfo{year}{2014}).


\bibitem[{\citenamefont{Pirandola}(2013{\natexlab{a}})\citenamefont{Pirandola}}]{Pirandola2014}
\bibinfo{author}{\bibfnamefont{S.}~\bibnamefont{Pirandola}},
\bibinfo{journal}{Sci. Rep.} \textbf{\bibinfo{volume}{5}},
  \bibinfo{pages}{6956} (\bibinfo{year}{2014}{\natexlab{a}}).

\bibitem[{\citenamefont{Ferraro et~al.}(2010)\citenamefont{Ferraro, Aolita,
  Cavalcanti, Cucchietti, and Acin}}]{Ferraro2010}
\bibinfo{author}{\bibfnamefont{A.}~\bibnamefont{Ferraro}},
  \bibinfo{author}{\bibfnamefont{L.}~\bibnamefont{Aolita}},
  \bibinfo{author}{\bibfnamefont{D.}~\bibnamefont{Cavalcanti}},
  \bibinfo{author}{\bibfnamefont{F.~M.} \bibnamefont{Cucchietti}},
  \bibnamefont{and} \bibinfo{author}{\bibfnamefont{A.}~\bibnamefont{Acin}},
  \bibinfo{journal}{Phys. Rev. A} \textbf{\bibinfo{volume}{81}},
  \bibinfo{pages}{052318} (\bibinfo{year}{2010}).

\bibitem[{\citenamefont{Fuchs and Van De~Graaf}(1999)}]{Fuchs1999}
\bibinfo{author}{\bibfnamefont{C.~A.} \bibnamefont{Fuchs}} \bibnamefont{and}
  \bibinfo{author}{\bibfnamefont{J.}~\bibnamefont{Van De~Graaf}},
  \bibinfo{journal}{Information Theory, IEEE Transactions on}
  \textbf{\bibinfo{volume}{45}}, \bibinfo{pages}{1216} (\bibinfo{year}{1999}).

\bibitem[{\citenamefont{de~Almeida et~al.}(2013)\citenamefont{de~Almeida, Gu,
  Fedrizzi, Broome, Ralph, and White}}]{de2013entanglement}
\bibinfo{author}{\bibfnamefont{M.}~\bibnamefont{de~Almeida}},
  \bibinfo{author}{\bibfnamefont{M.}~\bibnamefont{Gu}},
  \bibinfo{author}{\bibfnamefont{A.}~\bibnamefont{Fedrizzi}},
  \bibinfo{author}{\bibfnamefont{M.~A.} \bibnamefont{Broome}},
  \bibinfo{author}{\bibfnamefont{T.~C.} \bibnamefont{Ralph}}, \bibnamefont{and}
  \bibinfo{author}{\bibfnamefont{A.~G.} \bibnamefont{White}},
\bibinfo{journal}{Phys. Rev. A} \textbf{\bibinfo{volume}{89}},
  \bibinfo{pages}{042323} (\bibinfo{year}{2014}).

\bibitem[{\citenamefont{Shapiro}(2009)}]{Shapiro2009a}
\bibinfo{author}{\bibfnamefont{J.~H.} \bibnamefont{Shapiro}},
  \bibinfo{journal}{Phys. Rev. A} \textbf{\bibinfo{volume}{80}},
  \bibinfo{pages}{022320} (\bibinfo{year}{2009}).

\bibitem[{\citenamefont{Pirandola et~al.}(2008)\citenamefont{Pirandola,
  Mancini, Lloyd, and Braunstein}}]{Pirandola2008}
\bibinfo{author}{\bibfnamefont{S.}~\bibnamefont{Pirandola}},
  \bibinfo{author}{\bibfnamefont{S.}~\bibnamefont{Mancini}},
  \bibinfo{author}{\bibfnamefont{S.}~\bibnamefont{Lloyd}}, \bibnamefont{and}
  \bibinfo{author}{\bibfnamefont{S.~L.} \bibnamefont{Braunstein}},
  \bibinfo{journal}{Nature Physics} \textbf{\bibinfo{volume}{4}},
  \bibinfo{pages}{726} (\bibinfo{year}{2008}).


\bibitem[{\citenamefont{Pirandola}(2011)}]{Pirandola2011}
\bibinfo{author}{\bibfnamefont{S.}~\bibnamefont{Pirandola}},
  \bibinfo{journal}{Phys. Rev. Lett.} \textbf{\bibinfo{volume}{106}},
  \bibinfo{pages}{090504} (\bibinfo{year}{2011}).

\bibitem[{\citenamefont{Pirandola et~al.}(2011)\citenamefont{Pirandola, Lupo,
  Giovannetti, Mancini, and Braunstein}}]{pirandola2011quantum}
\bibinfo{author}{\bibfnamefont{S.}~\bibnamefont{Pirandola}},
  \bibinfo{author}{\bibfnamefont{C.}~\bibnamefont{Lupo}},
  \bibinfo{author}{\bibfnamefont{V.}~\bibnamefont{Giovannetti}},
  \bibinfo{author}{\bibfnamefont{S.}~\bibnamefont{Mancini}}, \bibnamefont{and}
  \bibinfo{author}{\bibfnamefont{S.~L.} \bibnamefont{Braunstein}},
  \bibinfo{journal}{New Journal of Physics} \textbf{\bibinfo{volume}{13}},
  \bibinfo{pages}{113012} (\bibinfo{year}{2011}).

\bibitem[{\citenamefont{Dall'Arno et~al.}(2011)\citenamefont{Dall'Arno, Michele}}]{dallarno2012}
\bibinfo{author}{\bibfnamefont{M.}~\bibnamefont{Dall'arno}},
  \bibinfo{author}{\bibfnamefont{A.}~\bibnamefont{Bisio}},
  \bibinfo{author}{\bibfnamefont{G.}~\bibnamefont{D'ariano}},
  \bibinfo{journal}{New Journal of Physics} \textbf{\bibinfo{volume}{10}},
  \bibinfo{pages}{1241010} (\bibinfo{year}{2012}).

\bibitem[{\citenamefont{Modi et~al.}(2011)\citenamefont{Modi, Cable,
  Williamson, and Vedral}}]{Modi2011a}
\bibinfo{author}{\bibfnamefont{K.}~\bibnamefont{Modi}},
  \bibinfo{author}{\bibfnamefont{H.}~\bibnamefont{Cable}},
  \bibinfo{author}{\bibfnamefont{M.}~\bibnamefont{Williamson}},
  \bibnamefont{and} \bibinfo{author}{\bibfnamefont{V.}~\bibnamefont{Vedral}},
  \bibinfo{journal}{Phys. Rev. X} \textbf{\bibinfo{volume}{1}},
  \bibinfo{pages}{021022} (\bibinfo{year}{2011}).



\bibitem[{\citenamefont{Regy et~al.}(2011)\citenamefont{Ragy, Berchera,
  Degiovanni, Olivares, Paris, Adesso and Genovese}}]{ragy2014}
\bibinfo{author}{\bibfnamefont{S.}~\bibnamefont{Ragy}},
  \bibinfo{author}{\bibfnamefont{I.}~\bibnamefont{Berchera}},
  \bibinfo{author}{\bibfnamefont{I.}~\bibnamefont{Degiovanni}},
    \bibinfo{author}{\bibfnamefont{S.}~\bibnamefont{Olivares}},
    \bibinfo{author}{\bibfnamefont{M.}~\bibnamefont{Paris}},
       \bibinfo{author}{\bibfnamefont{G.}~\bibnamefont{Adesso}}
  \bibnamefont{and} \bibinfo{author}{\bibfnamefont{M.}~\bibnamefont{Genovese}},
  \bibinfo{journal}{JOSA B} \textbf{\bibinfo{volume}{31}},
  \bibinfo{pages}{2045-2050} (\bibinfo{year}{2014}).


\bibitem[{\citenamefont{Invernizzi et~al.}(2011)\citenamefont{Invernizzi,
  Paris, and Pirandola}}]{Invernizzi2011}
\bibinfo{author}{\bibfnamefont{C.}~\bibnamefont{Invernizzi}},
  \bibinfo{author}{\bibfnamefont{M.~G.~A.} \bibnamefont{Paris}},
  \bibnamefont{and}
  \bibinfo{author}{\bibfnamefont{S.}~\bibnamefont{Pirandola}},
  \bibinfo{journal}{Phys. Rev. A} \textbf{\bibinfo{volume}{84}},
  \bibinfo{pages}{022334} (\bibinfo{year}{2011}).

\end{thebibliography}

\appendix

\section{Proof of Supporting Statements.} This section provides detailed proof of the two supporting statements. Recall that we introduced a variant of quantum illumination; where Alice's choice how to estimate $x$ when given $\rhocode[x]_{AB}$ is constrained: Alice is required to first make a local measurement on the idler $B$, followed by a local measurement on the signal $A$. Let $I'_c$ be the the optimal performance of this strategy. That is, Alice uses the above strategy to obtain $X_{est}$, an estimate of $X$. Let $I'_C$ be be the maximum $I(X_{est}, X)$ that can be achieved using the above strategy.

Here we prove the two supporting statements:
\begin{itemize}
\item[(i)] The optimal performance of quantum illumination, subject to the restriction of measure the idler first, followed by measurement of the signal, coincides with the best performance using conventional illumination $I'_c = \imaxc$.
\item[(ii)] The loss in performance in using this third approach over quantum illumination is $I_q - I'_c = \delta_{\mathrm{enc}}(A|B)$.
\end{itemize}
The detailed proofs of (i) and (ii) are below. Together, they imply that the main result, i.e., $I_q - \imaxc = \delta_{\mathrm{enc}}(A|B)$.

\textbf{Proof of Statement (i).} Recall that in the conventional
approach, Alice probes for the reflecting object with a pure qudit
state $\Phi=\ket{\phi}\bra{\phi}$. The target-variable $X$ is then
mapped into the output codewords
\begin{equation} \rhocode[0]_c = \eta \Phi+ (1 - \eta) \rho_E,
\qquad \rhocode[1]_c = \rho_E,
\end{equation}
with associated ensemble $\varepsilon_c = \{p_x, \rhocode[x]_c\}$. Alice's optimal performance $I_c$ is then given by, $I_{acc}(\varepsilon_c)$, the accessible information about $X$ with respect to the ensemble $\varepsilon_C$. Since the two codewords commute, this is equal to the Holevo information of communicating $x$ using $\rhocode[x]_c$ as codewords. That is, Alice's performance for a particular probe $\Phi$ is
\begin{equation}I_c (\Phi) = I_{acc}(\varepsilon_c) = S(\bar{\rho_c}) - \sum_x p_x S[\rho_c^{(x)}],
\end{equation}
where $\bar{\rho_c}=\sum_x p_x \rho_c^{(x)}$ is the
output state averaged over codewords, and $S(\cdot)$ the von Neumann entropy. The optimal conventional performance is given by the optimization
$\imaxc=\max_{\Phi} I_c (\Phi)$ over all possible pure states
$\Phi$.

Note that we are restricting such an optimization to pure states,
since mixed states surely provide worse performance. This can be
explicitly proven by \textit{reductio ad absurdum}. Assume that
there exists some mixed state $\rho = \sum_j \lambda_j
\ket{\phi_j}\bra{\phi_j}$ such that $I_c(\rho) > I_c(\Phi)$ for
all pure $\Phi$. Then, let $\sum_j \sqrt{\lambda_j}
\ket{\phi_j}_c\ket{j}_r$ be a purification of $\rho$, where $r$
denotes a reference system. If we had access to $r$, we can
measure it in the $\ket{j}$ basis. This would collapse the probe
state to $\ket{\phi_j}$ with probability $\lambda_j$, resulting in
an average performance of $\sum \lambda_j I_c
(\ket{\phi_j}\bra{\phi_j})$. In comparison, if the measurement
result was lost, our performance would reduce to $I_c (\rho)$.
Clearly, since performance can only degrade upon loss of
information
\begin{equation}
I_c (\rho) \leq \sum \lambda_j I_c (\ket{\phi_j}\bra{\phi_j}) \leq
\max_j I_c (\ket{\phi_j}\bra{\phi_j}).
\end{equation}
Therefore, there is a pure state $\ket{\phi_j}\bra{\phi_j}$ for
some $j$ such that $I_c (\rho) < I_c(\ket{\phi_j}\bra{\phi_j})$,
which contradicts our initial assumption. Hence $I_c$ must attain
its maximum on a pure state.

It is now important to note that, since $\rho_E$ is completely
mixed, symmetry considerations imply that $I_c (\Phi)$ is the same
for any pure state $\Phi$, i.e., all pure probes deliver equal
performance, and this performance coincides with the best possible
performance of conventional illumination $\imaxc$. This is a
simple consequence of the invariance of the Holevo information
under unitaries. In fact, let us apply an arbitrary unitary $U$ to
the codewords $\rho_c^{(x)}$ just before detection. Since $\rho_E$
is proportional to the identity, we have
\begin{equation} \tilde{\rho}_c^{(0)}:= U \rho_c^{(0)} U^{\dagger}=
\eta U \Phi U^{\dagger}+ (1 - \eta) \rho_E, \label{Holapp}
\end{equation}
and $\tilde{\rho}_c^{(1)}:= U \rho_c^{(1)} U^{\dagger}= \rho_E$.
These two codewords can equivalently be generated if we had
started from the input state $U \Phi U^{\dagger}$, which spans all
the Hilbert space by varying $U$. At the same time, we note that
the Holevo information does not change, i.e., for any $U$ we have
\begin{equation}
 I_c (U \Phi U^{\dagger})=\chi(\tilde{\rho}_c^{(0)},\tilde{\rho}_c^{(1)})
=\chi(\rhocode[0]_c,\rhocode[1]_c)= I_c (\Phi).
\end{equation}
Thus, find that $\imaxc=I_c (\Phi)$ for an arbitrary pure state
$\Phi$.

To demonstrate that the optimal conventional performance $\imaxc$
coincides with $I'_c$, we observe that all operations in the
quantum illumination circuit commutes with a local measurement on
the idler system $B$. Thus, there is no functional difference
between measuring the idler beam after receipt of the reflected
signal and measuring the same idler beam prior to sending out the
signal.

Then, suppose that Alice detects the idler system before
transmission, by applying a rank-1 POVM $\{\Pi_b\}$ on the
$B$-part of the maximally entangled state $\Psi_{AB}$. Given an
outcome $b$, with probability $q_b$, the signal system $A$ is
collapsed into a conditional pure state
$\Psi_{A|b}=q_b^{-1}\mathrm{Tr}_{B}(\Psi_{AB}\Pi_b)$ (this because
rank-1 POVMs project pure states into pure states). Sending any
pure probe $\Psi_{A|b}$ attains the maximum conventional
performance $\imaxc=I_c (\Psi_{A|b})$. On average, the performance
of Alice is therefore given by
\begin{equation}
I'_c = \sum_b q_b I_c (\Psi_{A|b})=\imaxc.\label{resultAPP1}
\end{equation}


\textbf{Proof of Statement (ii).} Suppose that Alice performs the
quantum illumination protocol by probing the target with the
$A$-part of a maximally entangled state $\Psi_{AB}$. The
target-variable $X=\{x,p_{x}\}$ is then mapped into the codewords
\begin{equation}
\rho_{AB}^{(0)}=\eta\Psi_{AB}+(1-\eta)(\rho_{E}\otimes\rho_{B}),~~\rho_{AB}^{(1)}=\rho_{E}\otimes\rho_{B}.
\end{equation}
with associated emsemble $\varepsilon_q = \{p_x, \rhocode[x]_{AB}\}$.
Let be the maximum amount of information that Alice retrieves about the
target-variable $X$ using arbitrary quantum measurements, then $I_{q} = I_{acc}(\varepsilon_q)$, the accessible information about $X$ with respect to $\varepsilon_q$. It is easy to check that, for $\Psi_{AB}$ maximally-entangled and $\rho_{E}$ maximally-mixed, $[\rho_{AB}
^{(0)},\rho_{AB}^{(1)}]=0$. Thus, $I_{acc}(\varepsilon_q)$ is equal to the Holevo information, i.e.,
\begin{align}
I_{q} &  = S(\bar{\rho}_{AB}) - \sum_x p_x S\left(\rhocode[x]_{AB}\right)\nonumber\\
&
=S(\bar{\rho}_{AB})-p_{0}S(\rho_{AB})-p_{1}[S(\rho_{E})+S(\rho_{B})],
\end{align}
where $\rho_{AB}=\rho_{AB}^{(0)}$ and
$\bar{\rho}_{AB}=p_{0}\rho_{AB} +p_{1}(\rho_{E}\otimes\rho_{B})$.

As before, let us consider $I^{\prime}_{c}$, defined as the
maximum accessible information on $X$ when Alice is constrained to
measure the idler before sending the signal. Here we prove that
\begin{equation}
I_{q} = I^{\prime}_{c} + \delta_{\mathrm{enc}}(A|B).
\end{equation}

In order to explicitly evaluate $I_{c}^{\prime}$, we apply
Eq.~(\ref{resultAPP1}), which can equivalently be written as
\begin{equation}
I_{c}^{\prime}=\sup_{\{\Pi_{b}\}}\sum_{b}q_{b}I_{c}(\Psi_{A|b}),
\label{resultAPP2}
\end{equation}
since any local rank-1 POVM $\{\Pi_{b}\}$ is optimal. Here, we
have
\begin{equation}
I_{c}(\Psi_{A|b})=\mathrm{SD}(\rho_{A|b}^{(0)},\rho_{A|b}^{(1)}),
\end{equation}
where
\begin{equation}
\rho_{A|b}^{(x)}=q_{b}^{-1}\mathrm{Tr}_{B}(\rho_{AB}^{(x)}\Pi_{b}).
\end{equation}
Since the two conditional codewords
\begin{equation}
\rho_{A|b}^{(0)}:=\rho_{A|b}=\eta\Psi_{A|b}+(1-\eta)\rho_{E},
\end{equation}
and $\rho_{A|b}^{(1)}=\rho_{E}$ commute, we can resort to the
Holevo information and write
\begin{equation}
I_{c}(\Psi_{A|b})=S(\bar{\rho}_{A|b})-p_{0}S(\rho_{A|b})-p_{1}S(\rho_{E}),
\label{IcAPP}
\end{equation}
where
\begin{equation}
\bar{\rho}_{A|b}=p_{0}\rho_{A|b}+p_{1}\rho_{E}=p_{0}\eta\Psi_{A|b}%
+(1-p_{0}\eta)\rho_{E} \label{rohabAPP}
\end{equation}
is the conditional output state averaged on the presence or not of
the target. By using Eq.~(\ref{IcAPP}) into
Eq.~(\ref{resultAPP2}), we get
\begin{equation}
I_{c}^{\prime}=\sup_{\{\Pi_{b}\}}\Big\{\sum_{b}q_{b}\Big[S(\bar{\rho}%
_{A|b})-p_{0}S(\rho_{A|b})\Big]\Big\}-p_{1}S(\rho_{E}). \label{IcpAPP}%
\end{equation}

Now it is important to note that, in the previous equation, the
von Neumann entropies $S(\bar{\rho}_{A|b})$ and $S(\rho_{A|b})$ do
not depend on the pure state $\Psi_{A|b}$. In fact, for any pure
$\Psi_{A|b}$, we can expand
the environmental state $\rho_{E}$ as%
\begin{equation}
\rho_{E}=d^{-1}\mathbb{I}=d^{-1}\left(
\Psi_{A|b}+\sum_{i=1}^{d-1}\left\vert i\right\rangle \left\langle
i\right\vert \right)  ,
\end{equation}
where $\left\langle i\right\vert \Psi_{A|b}\left\vert
i\right\rangle =0$ for any $i$. By replacing this expansion in
Eq.~(\ref{rohabAPP}), we find the spectral decomposition
\begin{equation}
\bar{\rho}_{A|b}=\lambda\Psi_{A|b}+\lambda_{\perp}\sum_{i=1}^{d-1}\left\vert
i\right\rangle \left\langle i\right\vert ~,\label{expAPP}
\end{equation}
with probabilities
\begin{equation}
\lambda:=p_{0}\eta+\frac{1-p_{0}\eta}{d},~\lambda_{\perp}:=\frac{1-p_{0}\eta
}{d}=\frac{1-\lambda}{d-1},\label{lambda2}
\end{equation}
where $\lambda_{\perp}$ is $(d-1)$ degenerate. The von Neumann
entropy $S(\bar{\rho}_{A|b})$ is equal to the Shannon entropy
associated with the previous probability distribution, i.e.,
\begin{equation}
S(\bar{\rho}_{A|b})=-\lambda\log_{2}\lambda-(1-\lambda)\log_{2}\frac
{1-\lambda}{d-1}.
\end{equation}
It is clear that the spectral decomposition of Eq.~(\ref{expAPP})
is exactly the same whatever the pure state $\Psi_{A|b}$ is. Thus,
its entropy $S(\bar{\rho }_{A|b})$ is independent from the
specific pure state $\Psi_{A|b}$ selected by the measurement
operator of the rank-1 POVM. The reasoning can be repeated for the
other state $\rho_{A|b}$, which has the same spectral
decomposition of $\bar{\rho}_{A|b}$ proviso that we set $p_{0}=1$
in Eq.~(\ref{lambda2}).

Therefore we have that $S(\bar{\rho}_{A|b})$ and $S(\rho_{A|b})$,
and also their difference
$\Delta_{b}:=S(\bar{\rho}_{A|b})-p_{0}S(\rho_{A|b})$, do not
depend on the pure state $\Psi_{A|b}$: These quantities are the
same for any choice of the measurement operator $\Pi_{b}$ of any
rank-1 POVM $\{\Pi_{b}\}$. As a result of this
measurement-independence, we can pick an arbitrary outcome
$\tilde{b}$ of an arbitrarily chosen rank-1 POVM and write the
following \begin{equation}
\sup_{\{\Pi_{b}\}}\sum_{b}q_{b}\Delta_{b}=\Delta_{\tilde{b}}.
\label{sup1}
\end{equation}
Because of the measurement-independence, we can also write
\begin{align}
S(\bar{\rho}_{A|\tilde{b}})  &
=\inf_{\{\Pi_{b}\}}\sum_{b}q_{b}S(\bar{\rho
}_{A|b}):=\bar{S}_{\min}(A|B),\\
S(\rho_{A|\tilde{b}})  &  =\inf_{\{\Pi_{b}\}}\sum_{b}q_{b}S(\rho
_{A|b}):=S_{\min}(A|B),
\end{align}
so that we find
\begin{equation}
\Delta_{\tilde{b}}=\bar{S}_{\min}(A|B)-p_{0}S_{\min}(A|B).
\label{delta1}
\end{equation}
By using Eqs.~(\ref{sup1}) and~(\ref{delta1}) in
Eq.~(\ref{IcpAPP}) we then write \begin{equation}
I_{c}^{\prime}=\bar{S}_{\min}(A|B)-p_{0}S_{\min}(A|B)-p_{1}S(\rho_{E}).
\end{equation}

Now we are ready compute the difference $I_{q}-I_{c}^{\prime}$,
which is given by
\begin{equation}
I_{q}-I_{c}^{\prime}=p_{0}\delta(A|B)-[S(\rho_{B})-S(\bar{\rho}_{AB})+\bar
{S}_{\min}(A|B)], \label{final0}
\end{equation}
where
\begin{equation}
\delta(A|B)=S(\rho_{B})-S(\rho_{AB})+S_{\min}(A|B)
\end{equation}
can be recognized to be the discord of $\rho_{AB}$. Note that
$\rho_{B} ^{(x)}=\mathrm{Tr}_{A}[\rho_{AB}^{(x)}]$ is equal to
$\rho_{B}=\mathrm{Tr} _{A}(\Psi_{AB})$ for any $x$, so that
\begin{equation}
\bar{\rho}_{B}:=p_{0}\rho_{B}^{(0)}+p_{1}\rho_{B}^{(1)}=\rho_{B}.
\end{equation}
This means that we can use the equality
$S(\rho_{B})=S(\bar{\rho}_{B})$ in Eq.~(\ref{final0}), which gives
\begin{align}
I_{q}-I_{c}^{\prime}  &
=p_{0}\delta(A|B)-[S(\bar{\rho}_{B})-S(\bar{\rho
}_{AB})+\bar{S}_{\min}(A|B)]\nonumber\\
&  =p_{0}\delta(A|B)-\bar{\delta}(A|B).
\end{align}
where $\bar{\delta}(A|B)$ is the discord of the average state
$\bar{\rho} _{AB}$. Thus, we finally get
$I_{q}-I_{c}^{\prime}=\delta_{\mathrm{enc}}(A|B)$ proving our
statement (ii). Combining this with statement (i) proves the main
result of our manuscript, i.e., $\Delta
I=\delta_{\mathrm{enc}}(A|B)$.

\end{document}